\documentclass[twocolumn,showpacs,preprintnumbers,amsmath,amssymb]{revtex4}
\usepackage{graphicx}% Include figure files
\usepackage{dcolumn}% Align table columns on decimal point
\usepackage{bm}% bold math

\def\beq{\begin{equation}}
\def\eeq{\end{equation}}
\def\bw{\begin{widetext}}
\def\ew{\end{widetext}}
\def\pl{\partial}

\def\al{\alpha}
\def\bt{\beta}

\def\de{\delta}

\def\Si{\Sigma}
\def\te{\theta}
\def\La{\Lambda}
\def\lam{\lambda}
\def\Om{\Omega}

\def\sq{\sqrt}

\def\l{\left (}
\def\r{\right )}
\def\fr{\frac}
\def\la{\label}
\def\hs{\hspace}
\def\vs{\vspace}
\def\inf{\infty}
\def\ran{\rangle}
\def\lan{\langle}
\def\ov{\overline}
\def\tl{\tilde}
\def\tm{\times}

\begin{document}

%\preprint{APS/123-QED}

\vspace{0.5cm} 

\title{SU(4)$_{\rm c}\times $SU(2)$_{\rm L}\times $SU(2)$_{\rm R}$ model\\
{}from 5D SUSY SU(4)$_{\rm c}\times $ SU(4)$_{\rm L+R}$ }

\author{Qaisar Shafi$^a$ and
Zurab Tavartkiladze$^{b, c}$ 
}

\address{{\it $^a$ Bartol Research Institute, University of Delaware,
Newark, DE 19716, USA\\
$^b$Institute for Theoretical Physics, Heidelberg University,
Philosophenweg 16, D-69120 Heidelberg, Germany\\
$^c$Institute of Physics, Georgian Academy of Sciences, Tbilisi 380077,
Georgia }}

\begin{abstract}
\vspace{0.3cm}

We investigate supersymmetric $SU(4)_c\times SU(4)_{L+R}$ theory in
5 dimensions whose compactification on a $S^{(1)}/Z_2$ orbifold yields 
$N=1$ supersymmetric $SU(4)_c\times SU(2)_L\times SU(2)_R$ supplemented by
a $\tl{U}(1)$ gauge symmetry. We discuss how the $\mu $
problem is resolved,
a realistic Yukawa sector achieved, and a stable proton realized.
Neutrino
masses and oscillations are also briefly discussed. 

\end{abstract}

% PACS, the Physics and
%Astronomy
\maketitle

\section{Introduction}

The minimal supersymmetric standard model (MSSM) provides an elegant,
albeit only partial resolution of the gauge hierarchy problem. The
apparent unification at $M_G$($\sim 2\cdot 10^{16}$~GeV) of the three
MSSM gauge couplings \cite{unif}
hints at the existence of an underlying supersymmetric grand unified
theory (SUSY GUT). Recent excitement about higher dimensional SUSY GUTs
is primarily sparked by the observation \cite{dtpdecay} that the notorious
doublet-triplet (DT) splitting problem can be resolved through suitable
boundary conditions on an appropriately chosen orbifold \cite{dtpdecay},
\cite{symbr}. 
%Any possible mismatch
%between $M_G$ and $M_P$ can be understood by appropriately adjusting
%the size of the extra dimension(s).
The magnitude of $M_G$ also seems consistent
with the scale of lepton number violation, thereby providing the
possibility
of explaining the atmospheric \cite{atm} and solar \cite{sol} neutrino
anomalies through neutrino oscillations.
 
In the absence of extra dimensions, the choice of SUSY GUT is normally
dictated by the requirement that the known matter multiplets fall into
chiral families \cite{chiran}. This imposes limitations on possible gauge
groups, and in 4D, the simplest choices are $SU(5)$, $SO(10)$, etc.
However, in the presence of extra dimensions, dimensional reduction can
yield chiral 'zero' modes on
a boundary subspace, even if the initial 5 (or higher) dimensional theory
is vector-like. This opens up the possibility of discussing new unified
groups in higher dimensions.

In this paper we follow this reasoning and consider SUSY 
$SU(4)_c\times SU(4)_{L+R}$
gauge theory [which, in turn, can be embedded say in $SO(12)$] in five   
dimensions. Through compactification on a $S^{(1)}/ Z_2$ orbifold, we
obtain  four dimensional 
$N=1$ SUSY $SU(4)_c\times SU(2)_L\times SU(2)_R$ \cite{PS}, supplemented
by an apparently anomalous $\tl{U}(1)$ symmetry. The desired 'matter' as
well
as 'scalar' supermultiplets emerge after proper selection of parities
under $Z_2$. At the 4D level, the anomalies due to $\tl{U}(1)$ are
canceled by the
contribution from bulk Chern-Simons term \cite{5dact}-\cite{bl}
and additional states on fixed point(s) are crucial.
Let us note that there are wide class of string constructing models, which
in 4D low energetical limit give anomalous $U(1)$ factors. The latter 
can be
canceled by the Green-Schwarz mechanism \cite{GS}. The 5D scenario
considered
here is transparent field-theoretical example of GUT model, which after
compactification breaking gives anomalous $\tilde U(1)$.
Let us note also that in \cite{su3w}, \cite{1su3w} the 5D $SU(3)_W$
unified models, involving gauge symmetry and 
SUSY breaking, were considered and got discussed extensively 
consistency of these type of models  from the viewpoint of anomaly
cancellation \cite{locan}-\cite{bl}. 
Together with Chern-Simons term, the selected additional states
on the fixed points can play crucial role for anomaly
cancellation \cite{1su3w}, \cite{bl}.
In difference from $SU(3)_W$ model,
considered $SU(4)_c\tm SU(4)_{L+R}$ scenario unifies left-right 
$SU(2)_L\tm SU(2)_R$ group (of Pati-Salam \cite{PS}) in a single
$SU(4)_{L+R}$. Being broken at the low energies, the latter could give
interesting phenomenological implications.

We discuss how realistic phenomenology emerges in 4D. An extra triplet      
state decouples at a high mass scale, while the DT splitting problem does    
not exist at all. The breaking of $\tl{U}(1)$ symmetry is guaranteed by
the Fayet-Iliopoulos D-term, with the consequence that a realistic 
Yukawa sector can  
be realized. After reduction of the 5D theory, there are some 
${\cal R}$-symmetries in
4D which can provide for an automatic baryon number conservation,
thereby guaranteeing a stable proton.
The same ${\cal R}$-symmetry is also crucial for avoiding an unacceptably
large $\mu $ term in unbroken $N=1$ SUSY limit.
We also briefly consider neutrino masses as well as oscillations.

\section{5D SUSY $SU(4)_c\times SU(4)_{L+R}$ model 
and its orbifold breaking}

We consider a $SU(4)_c\times SU(4)_{L+R}\equiv G_{44}$ sypersymmetric model in
5D dimension. In 4D notation we have $N=2$ SUSY, where the chiral
supermultiplet
${\bf \Phi}_{N=2}=(\Phi, \ov{\Phi })$ contains two $N=1$ chiral
supermultiplets
$\Phi $, $\ov{\Phi }$ transforming as $p$ and $\ov{p}$-plets respectively
under the gauge group. Under $\Phi $ we denote all 'matter' and/or 'scalar'
superfields of the model, while $\ov{\Phi }$ indicate their mirrors.
The $N=2$ gauge superfield is
${\bf V}_{N=2}=(V, \Si )$, where $V$ and $\Si $ are $N=1$ vector and
chiral
superfields in the adjoint representation of the gauge group.
In terms of $N=1$ components, the 5D action includes
\cite{5dact}:

\beq
S^{(5)}=\int d^5x({\cal L}^{(5)}_V+{\cal L}^{(5)}_{\Phi })
\la{5dact},
\eeq
where

$$
{\cal L}^{(5)}_V=
\fr{1}{g^2}\int d^4\te \l (\sq{2}\pl_5V+\Si^{+})e^{-V}
(-\sq{2}\pl_5V+\Si )e^V\right.+
$$
\beq
\left. \pl_5e^{-V}\pl_5e^V\r+
\fr{1}{4g^2}\int d^2\te W^{\al }W_{\al }+{\rm H.c.}~,
\la{lagv}
\eeq

$$
{\cal L}^{(5)}_{\Phi }=\int d^4\te \l \Phi^{+}e^{-V}\Phi +
\ov {\Phi }e^{V}\ov {\Phi }^{+} \r+
$$
\beq
\int d^2\te \ov{\Phi}\l M_{\Phi }+\pl_5 -\fr{1}{\sq{2}}\Si \r \Phi+
{\rm H.c.},
\la{lagf}
\eeq
and $W_{\alpha}$ are the supersymmetric field strengths.
The action in (\ref{5dact}) is invariant under the gauge
transformations:
$$
e^V\to e^{\La }e^Ve^{\La^{+} }~,~~~~
\Si \to e^{\La }(\Si-\sq{2}\pl_5)e^{-\La }~,
$$
\beq
\Phi \to e^{\La }\Phi~,~~~~~~~~\ov{\Phi } \to \ov{\Phi }e^{-\La }~.
\la{n2gaugetr}
\eeq

%\subsection{Matter and gauge sector}

Next we introduce one matter $N=2$
supermultiplet ${\bf F}_{N=2}=({\bf F}, {\bf \ov{F}})$ per generation
where, under $G_{44}$, ${\bf F}$ and ${\bf \ov{F}}$ transform as

\beq
{\bf F}\sim (4, 4)\equiv (\ov {F}^{\hs{1mm}c}, ~{F} )~,~~~~~
\ov{\bf F}\sim (\bar 4, \bar 4)\equiv ({F}^{\hs{0.5mm}c},~\ov {F})~,
\la{matter}   
\eeq
with

$$
\begin{array}{cc}
\ov {F} ^{\hs{1mm}c}\hs{-1mm}=\!\!\!\!\hs{-1mm} &
\left (\begin{array}{ccc}
\hs{-1mm}\ov u_1^{\hs{1mm}c}~,
&\ov d_1^{\hs{1mm}c}
\\
\hs{-1mm}\ov u_2^{\hs{1mm}c}~,
&\ov d_2^{\hs{1mm}c}
\\
\hs{-1mm}\ov u_3^{\hs{1mm}c}~,
&\ov d_3^{\hs{1mm}c}
\\
\hs{-1mm}\ov \nu^{\hs{1mm}c}~,
&\ov e^{\hs{1mm}c}
\end{array}\hs{-1.5mm}\right)\hs{0.1mm},~~~~
\end{array}
\begin{array}{cc}
{F}\hs{-1mm}=\!\!\!\!\!\hs{-1mm} &{\left(\begin{array}{ccc}
\hs{-1mm} u_1~, &d_1
\\
\hs{-1mm} u_2~, &d_2
\\
\hs{-1mm} u_3~, &d_3
\\
\hs{-1mm} \nu ~, &e
\end{array}\hs{-1.5mm}\right)\hs{0.1mm},
}
\end{array}
%%%%%%%%%%%%
$$
\beq
\begin{array}{cc}
{F}^{\hs{0.5mm}c}\hs{-1mm}=\!\!\!\!\hs{-1mm} & \left
(\begin{array}{ccc}
\hs{-1mm}u_1^{\hs{1mm}c}~, &d_1^{\hs{1mm}c}
\\
\hs{-1mm} u_2^{\hs{1mm}c}~, & d_2^{\hs{1mm}c}
\\
\hs{-1mm} u_3^{\hs{1mm}c}~, & d_3^{\hs{1mm}c}
\\
\hs{-1mm} \nu^{\hs{1mm}c}~, & e^{\hs{1mm}c}
\end{array}\hs{-1.5mm}\right)\hs{0.1mm},~~~~
\end{array}
\begin{array}{cc}
\ov{F}\hs{-1mm}=\!\!\!\!\!\hs{-1mm} &{\left(\begin{array}{ccc}
\hs{-1mm} \ov u_1~,  &\ov d_1
\\
\hs{-1mm} \ov u_2~, &\ov d_2
\\
\hs{-1mm} \ov u_3~, &\ov d_3
\\
\hs{-1mm} \ov{\nu }  ~, &\ov e
\end{array}\hs{-1.5mm}\right)\hs{0.1mm}.
}
\end{array}
\label{Fs}
\end{equation}
It is clear that (\ref{Fs}) is just the chiral content of 
$SU(4)_c\times SU(2)_L\times SU(2)_R\equiv G_{422}$ model \cite{PS},
supplemented by mirrors. It will turn out that mirrors can be projected out
after compactification of the fifth dimension on the orbifold $S^{(1)}/Z_2$. 
The $G_{44}$ symmetry is broken by orbifolding via the channel

\beq
SU(4)_c\times SU(4)_{L+R}\to SU(4)_c\times SU(2)_L\times SU(2)_R
\times \tl{U} (1)~,
\la{4221}
\eeq
where the generator of $\tl{U} (1)$ is

\beq
Y_{\tl{U}(1)}=\fr{1}{\sq{8}}\cdot {\rm Diag}\l 1,~ 1,~ -1,~ -1 \r~,
\la{u1ch}
\eeq
with $SU(4)_{L+R}$ normalization.
In terms of $G_{422}\times \tl{U}(1)$, the fermionic fragments read

$$
\ov{F}^c=(4,~ 1,~ 2)_{-1}~,~~~~~F=(4,~ 2,~ 1)_1~,
$$
\beq
F^c=(\bar 4, ~1, ~\bar 2 )_1~,~~~~\ov{F}=(\bar 4,~ \bar 2,~ 1)_{-1}~,
\la{decmat}
\eeq
where subscripts denote $\tl{U} (1)$ charges in units of $1/\sq{8}$
[see (\ref{u1ch})].
Decomposition of $SU(4)_{L+R}$ adjoint yields

\beq
15_{L+R}=(3,~ 1)_0+(1,~ 3)_0+(1,~ 1)_0+
(\bar 2,~ 2)_{-2}+(2,~ \bar 2)_2~.
\la{decgauge}
\eeq

%\beq
%(\ov {F}^{\hs{1mm}c},~ \ov {F})\to
%-(\ov {F}^{\hs{1mm}c},~\ov {F})~,~~~~~
%({F},~ {F}^{\hs{1mm}c})\to
%({F},~ {F}^{\hs{0.5mm}c})~,
%\la{transf}
%\eeq
%
%
%$$
%\left [ V(3, 1),~ V(1, 3),~ V(1, 1)\right ]\to 
%\left [V(3, 1),~ V(1, 3),~ V(1, 1) \right ]~,
%$$
%$$
%\left [V(\bar 2, 2),~ V(2, \bar 2) \right ]\to
%-\left [ V(\bar 2, 2),~ V(2, \bar 2)\right ]~,
%$$
%$$
%\left [ \Si (3, 1),~ \Si (1, 3),~ \Si (1, 1)\right ]\to
%-\left [\Si (3, 1),~ \Si (1, 3),~ \Si (1, 1) \right ]~,
%$$
%$$
%\left [\Si (\bar 2, 2),~ \Si (2, \bar 2) \right ]\to
%\left [ \Si (\bar 2, 2),~ \Si (2, \bar 2)\right ]~,
%$$
%\beq
%V(15_c)\to V(15_c)~,~~~\Si (15_c)\to -\Si (15_c)~,
%\la{transvec}
%\eeq

Under $Z_2$ the fifth coordinate $y$ changes sign $y\to -y$.
Because of this, states with positive and negative parities $\Phi_{+}$,
$\ov{\Phi }_{-}$ can be written in factorized forms

$$
\Phi_{+}=\sum_{n=0}^{n=\inf }\Phi^{(n)}(x)
\cos \l \fr{ny}{R}\r~,
$$
\beq
\ov{\Phi }_{-}=\sum_{n=1}^{n=\inf }\ov{\Phi }^{(n)}(x)
\sin \l \fr{ny}{R}\r~.
\la{expan1}
\eeq
As expected, $\ov{\Phi }_{-}$ in (\ref{expan1}) lacks zero modes on
$S^{(1)}/Z_2$ orbifold. The fixed point $y=0$ will be identified with
a 3-brane corresponding to our 4D world.
The $Z_2$ parities  of the various fragments of the gauge 
(${\bf V}_{N=2}$)
and matter (${\bf F}_{N=2}$) superfields are shown in Table 1.

Under $Z_2$, the 
D-terms in (\ref{lagv}), (\ref{lagf}) involving
${\bf V}_{N=2}$ and ${\bf F}_{N=2}$ states
are invariant. The $F$-terms in
(\ref{lagf}) can be written schematically as 

%%%%%%%%%%%%%%%%%%%%%%%%%%%%%
% BEGIN TABLE
%%%%%%%%%%%%%%%%%%%%%%%%%%%%%

\begin{table}
\caption{$Z_2$ parity numbers of gauge, matter and scalar superfields }

\label{t:scalar}
$$\begin{array}{|c|c|c|}

\hline

&{\rm Multiplets~ in~ terms~ of} 
~G_{422}\times \tl{U}(1) &
\hs{-0.1cm}Z_2 {\rm par.}\hs{-0.1cm} \\ 
%\hline %xui

%\hline

\hline

{\bf V}_{N=2}  &

$$\begin{array}{ccccccccc}
&&V(15_c)~,&&V(3, 1)_0~,&&V(1, 3)_0~,&&\\
&&V(1, 1)_0~,&&\Si (\bar 2, 2)_{-2}~, &&\Si (2, \bar 2)_2 &&\\
\hline
&&\Si (15_c)~,&&\Si (3, 1)_0~,&&\Si (1, 3)_0~,&&\\
&&\Si (1, 1)_0~, &&V(\bar 2, 2)_{-2}~, &&V(2, \bar 2)_2 &&\\
\end{array}$$

 &

$$\begin{array}{ccc} 
&+&\\
\hline
&-&\\
\end{array}$$
\\
%\hline

\hline
{\bf F }_{N=2}  &

$$\begin{array}{ccccccccc}
&&&F(4, 2, 1)_1~,&&~~~F^c(\bar 4, 1, \bar 2)_1&&&\\
\hline
&&&\ov{F}(\bar 4, \bar 2, 1)_{-1}~,&&~~~
\ov{F}^{\hs{1mm}c}(4, 1, 2)_{-1}&&&\\
\end{array}$$

 &

$$\begin{array}{ccc}
 &+&\\
\hline
&-&\\
\end{array}$$
\\
%\hline

%\hline

\hline
{\bf H }_{N=2}  &

$$\begin{array}{ccccccccc}
&&&H^c(\bar 4, 1, \bar 2)_1~,&&~~~\ov{H}^{\hs{1mm}c}(4, 1, 2)_{-1}&&&\\
\hline
&&&H(4, 2, 1)_1~,&&~~~\ov{H}(\bar 4, \bar 2, 1)_{-1}&&&\\
\end{array}$$

 &

$$\begin{array}{ccc}
&+&\\
\hline
&-&\\
\end{array}$$
\\
%\hline

%\hline

\hline
{\bf \Om }_{N=2}  &

$$\begin{array}{c}
D_1^{~}(6, 1, 1)_2~, D_2(6, 1, 1)_{-2}~,
P(\bar 6, \bar 2, \bar 2)_0\\
\hline
\ov{D}_1(6, 1, 1)_2~, \ov{D}_2(6, 1, 1)_{-2}~,
P'(6, 2, 2)_0\\ %xui
\end{array}$$

& 

$$\begin{array}{ccc}  
&+&\\
\hline
&-&\\
\end{array}$$
\\
%\hline

%\hline

\hline
{\bf \Psi }_{N=2}  &

$$\begin{array}{c}
S(1, 1, 1)_{-2}~, \ov{S}(1, 1, 1)_2~,h(1, 2, 2)_0\\
\hline
S\hs{0.5mm}'(1, 1, 1)_{-2}~, \ov{S}\hs{0.5mm}'(1, 1, 1)_2~,
h'(1, 2, 2)_0\\
\end{array}$$

 &

$$\begin{array}{ccc}
 &+&\\
\hline
&-&\\
\end{array}$$
\\
\hline

\end{array}$$

\end{table}

%xui
%%%%%%%%%%%%%%%%%%%%%%%%%%%%%%%%%%%%
%%% END TABLE
%%%%%%%%%%%%%%%%%%%%%%%%%%%%%%%%%%%%

$$
F^c\l \pl_5 -\hs{-1mm}
\fr{1}{\sq{2}}\left [ \Si (15_c)+\hs{-1mm}
\Si (1, 3)+\hs{-1mm}\Si (1, 1) \right ] \r \ov{F}^c+
$$
$$
\ov{F}\l \pl_5 -\hs{-1mm}
\fr{1}{\sq{2}}\left [ \Si (15_c)+\hs{-1mm}\Si (3, 1)
+\hs{-1mm}\Si (1, 1) \right ] \r F
$$
\beq
-\fr{1}{\sq{2}}\ov{F}\Si (2, \bar 2) \ov{F}^c
-\fr{1}{\sq{2}}F^c \Si (\bar2, 2)  F
+~{\rm H.c.}~.
\la{fterms}
\eeq
Note that bare mass terms are forbidden by $Z_2$ symmetry.

{}From Table 1 the gauge fields with positive $Z_2$ parity are 
just those of $G_{422}\times \tl{U}(1)$. 
The $\Si $ states in the same representations
have opposite parities, so that compactification gives 4D $N=1$ SUSY.
By the same token, 
on $S^{(1)}/Z_2$ orbifold we only have $F$,
${F}^{\hs{0.5mm}c}$ chiral states. From the
charges in (\ref{decmat}) we see that the $\tl{U}(1)$ symmetry is
anomalous at 4D level
(although in 5D the theory is vector-like and anomaly free).
The appearance of chirality and anomaly on the boundary of higher
dimensional theory is not surprising and was observed a long time ago
\cite{chernsim}, \cite{lazshafi}. 
In our orbifold scenario localized anomalies on orbifold fixed points
$y=0$ (brane identified with our 4D world)and $y=\pi R$ (hidden
brane) are generated and divergence of five dimensional current has form
\cite{locan}

\beq
D_AJ^A=\fr{A}{2}[\de (y)+\de (y-\pi R)]~,
\la{diverg}
\eeq
where $A\sim {\rm Tr}[T^a\{T^b,~T^c\}]$. 

We see that the anomalies on the fixed
points have the same sign and differ by a factor $2$ from 4D anomalies. To
cancel the anomalies in (\ref{diverg}) and make the theory selfconsistent,
the bulk
Chern-Simons term can be applayed 
\cite{5dact}-\cite{bl}.
Since the CS term on boundaries induce contributions
with opposite signs, it is impossible to simultaneously cancel the
anomalies on both fixed points unless some additional states are
involved \cite{1su3w}, \cite{bl}. We will show how to add
some states on $y=\pi R$ brane such that the anomalies on $y=0$ and
$y=\pi R$ will have opposite signs, i.e. the condition

\beq
\fr{A}{2}+A_{\rm br}=-\fr{A}{2}
\la{ancond}
\eeq
will be satisfied. 

Since $G_{44}$ is broken to 
$G_{422}\tm \tl{U}(1)$ on both fixed points, we will add 
on $y=\pi R$ brane states which
transform under the latter gauge group. Namely, selecting the states as
$N_6\tm (6, 1, 1)_{-2}+N_2\tm (1, \bar 2, 2)_{-2}+N_1\tm (1, 1, 1)_2$
(where 
$N_6$, $N_{1, 2}$ denote number of appropriate representations), one can
check that for $(N_6, N_2, N_1)=(3, 3, 24)$ we will have

$$
A_{\rm br}[SU(4)_c^2-\tl{U}(1)]=A_{\rm br}[SU(2)_L^2-\tl{U}(1)]=
$$
\beq
A_{\rm br}[SU(2)_R^2-\tl{U}(1)]=-6~,~~~
A_{\rm br}[\tl{U}(1)^3]=-48~.
\la{branom}
\eeq
Since, with $3\tm (F+F^c)$ states we have

$$ 
A[SU(4)_c^2-\tl{U}(1)]=A[SU(2)_L^2-\tl{U}(1)]=
$$
\beq
A[SU(2)_R^2-\tl{U}(1)]=6~,~~~
A[\tl{U}(1)^3]=48~,
\la{bulkanom}
\eeq
using (\ref{branom}), (\ref{bulkanom}) the relation (\ref{ancond}) is
indeed
satisfied, which is sufficient to cancel anomalies by
the bulk CS term. Of course for the latter to work, there are a variety of
possibilities. Namely, one can introduce additional states either on both
fixed points, or in the bulk, or simultaneously on the fixed points and in
the bulk
\cite{bl}. We have presented just one example to
demonstrate how the anomaly cancellation mechanism works out and the
theory becomes selfconsistent.

%The induced 4D anomalies are
%canceled through contribution of the bulk Chern-Simons term 
%\cite{chernsim}. For a supersymmetrized version of this effect see
%\cite{5dact}.

\vs{0.2cm}

For $G_{422}$ breaking we need additional scalar superfields. For this
purpose, and also for decoupling some unwanted colored triplets
\cite{ant}, in 5D
we introduce $N=2$ supermultiplets 
${\bf H}_{N=2}=({\bf H}, \ov{\bf H})$ and
${\bf \Om}_{N=2}=(\Om , \ov{\Om })$. Under $G_{44}$,

$$
{\bf H}=(4,~ 4)=(\ov{H}^{\hs{1mm}c},~ H)~,~~~~
\ov{\bf H}=(\bar 4,~ \bar 4)=(H^c,~ \ov{H})~,
$$
\beq
\Om =(6, 6)~,~~~~\ov{\Om }=(\bar 6, \bar 6).
\la{scalar1}
\eeq
In terms of $G_{422}\times \tl{U}(1)$ the ${\bf H}$, $\ov{\bf H}$
decompose as ${\bf F}$ and ${\bf \ov{F}}$ 
[see (\ref{matter}), (\ref{Fs}), (\ref{decmat})], 
while for $\Om $, $\ov{\Om }$ we have

$$
\Om =(6, 6)=D_1(6, 1, 1)_2+D_2(6, 1, 1)_{-2}+P'(6, 2, 2)_0~,
$$
\beq
\ov{\Om }=(\bar 6, \bar 6)=\ov{D}_1(\bar 6, 1, 1)_{-2}+
\ov{D}_2(\bar 6, 1, 1)_{2}+P(\bar 6, \bar 2, \bar 2)_0~.
\la{decOm}
\eeq
The transformation properties of fragments
{}from ${\bf H}_{N=2}$, ${\bf \Om }_{N=2}$ under $Z_2$ orbifold symmetry
are given in Table 1.
In 4D we are left with $H^c$, $\ov{H}^{\hs{1mm}c}$, $D_1$, $D_2$, and $P$,
with the remaining states projected out. The state $P$ is self conjugate
and neutral under anomalous $\tl{U}(1)$, and at 4D level it gains mass through
the superpotential coupling $MP^2$ ($M\sim M_G$) and decouples. The states
$(\nu^c+\ov{\nu }^{\hs{1mm}c})_H$ from $H^c$ and
$\ov{H}^{\hs{1mm}c}$ respectively, develop non zero VEVs ($\sim M_G$)
so that the symmetry $G_{422}$ is broken down to
$SU(3)_c\times SU(2)_L\times U(1)_Y\equiv G_{321}$. For correct symmetry
breaking we have to avoid the mass term $M_H\ov{H}^{\hs{1mm}c}H^c$ [which otherwise
would cause unacceptable SUSY breaking in 4D]. This is easily 
achieved by introducing
$\ov{Z}_2$ symmetry (not to be confused with $Z_2$ orbifold symmetry), under
which $H\to -H$, $\ov{H}^{\hs{1mm}c}\to -\ov{H}^{\hs{1mm}c}$
(e.g. ${\bf H}\to -{\bf H}$). All D-terms in (\ref{lagf})
involving  ${\bf H}_{N=2}$ and ${\bf \Om }_{N=2}$ states are invariant
under $\ov{Z}_2$ and $Z_2$.

During $G_{422}$ breaking, the states $u^c,~ e^c$ and
$\ov{u}^{\hs{1mm}c},~\ov{e}^{\hs{1mm}c} $ from
$H^c$ and $\ov{H}^{\hs{1mm}c}$ are absorbed by the appropriate gauge fields,
while $d^c+\ov{d}^{\hs{1mm}c}$ are still massless. However, they can acquire masses by
mixings with appropriate states from $D_{1, 2}$. The relevant 4D
superpotential couplings are

\beq   
\lam_1H^cH^cD_2+\lam_2\ov{H}^{\hs{1mm}c}\ov{H}^{\hs{1mm}c}D_1+M_DD_1D_2~,
\la{supHD4d}
\eeq
where $\lam_{1, 2}$ are order unity dimensionless couplings, and 
$M_D\sim M_G$. After substituting appropriate VEVs in (\ref{supHD4d}),
mass matrix for color triplets is given by

\begin{equation}
\begin{array}{ccc}
 & {\begin{array}{ccc}
\hspace{-5mm}~\ov d_{\ov H^c}^{\hs{1mm}c} & \,\,~~
\ov d_{D_1}^{\hs{1mm}c} &
\,\,~\ov d_{D_2}^{\hs{1mm}c}
  
\end{array}}\\ \vspace{2mm}
M_T= \begin{array}{c}
d_{H^c}^{\hs{1mm}c}\\ d_{D_1}^{\hs{1mm}c} \\d_{D_2}^{\hs{1mm}c}
 \end{array}\!\!\!\!\! &{\left(\begin{array}{ccc}
\,\,0~~  &\,\,0~~ &\,\,\lam_1V~~
\\
\,\,\lam_2V~~   &\,\,0~~  &\,\,M_D~~
\\
\,\,0~~ &\,\,~M_D~~ &\,\,0~~
\end{array}\right) }~,
\end{array}  \!\!  ~~~~~
\label{tripmas}
\end{equation}
where $V$ is the VEV of $H^c$, $\ov{H}^{\hs{1mm}c}$ along the $G_{321}$
singlet direction. This shows that the triplets from  $H^c$,
$\ov{H}^{\hs{1mm}c}$ gain the masses through mixings with the triplets of
$D_{1, 2}$.

\section{Yukawa Sector}

Turning to the Yukawa sector,
recall that from ${\bf V}_{N=2}$ supermultiplets, the $N=1$ chiral
superfields
$\Si(\bar 2, 2)_{-2}\equiv \Si_{-2}$ and $\Si(2, \bar 2)_2\equiv \Si_2$
have positive $Z_2$ parities (see Table 1) and are therefore not
projected out.
These states are bi-doublets and $\Si_{-2}$ has coupling with matter even
at 5D level [last term in (\ref{fterms})]:
  
\beq
FF^c\Si_{-2}~.
\la{yuk1}
\eeq
However, $\Si_{-2}$ 
cannot contain the MSSM higgs doublets
because it forms a massive state with
$\Si_{2}$,

\beq
M_{\Si }\Si_{-2}\Si_{2}~.
\la{decoupl}
\eeq

In order to build a realistic Yukawa sector we introduce 
a new state
${\bf \Psi}_{N=2}=(\Psi, \ov{\Psi })$, where

$$
\Psi=(1, 6)=h(1, 2, 2)_0+\ov{S}\hs{0.5mm}'(1, 1, 1)_2+
S\hs{0.5mm}'(1, 1, 1)_{-2}~,
$$
\beq
\ov{\Psi }=(1, \bar 6)=h'(1, \bar 2, \bar 2)_0+\ov{S}(1, 1, 1)_2+
S(1, 1, 1)_{-2}~.
\la{decPsi}
\eeq

With the transformation properties for the fragments of ${\bf \Psi }_{N=2}$
presented in Table 1, the states $h$, $\ov{S}$, $S$ are not projected out.
These states turn out to be crucial for realistic Yukawa sector.
The D-terms in (\ref{lagf}), involving fragments of ${\bf \Psi }_{N=2}$, are
invariant under orbifold symmetry, while the
relevant allowed F-terms are $\ov{S}\Si_{-2}h$ and $S\Si_{2}h$.
Combining these two terms with (\ref{decoupl}),
the relevant superpotential couplings are

\beq
\ov{S}\Si_{-2}h+S\Si_{2}h+M_{\Si }\Si_{-2}\Si_{2}~.
\la{mixdoubl}
\eeq

Since $\tl{U}(1)$ is anomalous, the Fayet-Iliopoulos term 
$\xi \int d^4\te V_{\tl{U}(1)}$ will be allowed
in 4D, and we assume that $\xi>0$. 
The singlet $S$ has negative $\tl{U}(1)$ charge, so it can
be used for its breaking. So, Cancelling the $\xi$-term, 
$\lan S\ran \sim \sq{\xi }$, while $\lan \ov{S}\ran =0$.
{}From all this and from (\ref{mixdoubl}), one can see that the light
bi-doublet 
$\tl{h}$ belongs with equal weights to both $\Si_{-2}$ and $h$ 
[if $\lan S\ran \sim M_{\Si }$].
Taking into account (\ref{yuk1}), the Yukawa coupling

\beq
FF^c\tl{h}~,
\la{yuk2}
\eeq
is generated which, however yields' 
degenerate masses for
up-down quarks and charged leptons. This drawback exists in all minimal
versions of $G_{422}$, and for its resolution some additional
mechanisms must be applied \cite{masses422}, \cite{our422}. We do not go
through the details of
this issue here and refer the reader to \cite{masses422}, \cite{our422},
where realistic
patterns of fermion masses and mixings are constructed.
Let us note that the anomalous $\tilde{U}(1)$ also can be exploited for
understanding and solving various puzzles  such as mechanism of SUSY 
breaking, suppression of FCNC baryon number conservation etc. in the
spirit of refs. \cite{gia}, \cite{FCNC}. 

\section{$\mu$-Term and Baryon Number Conservation}

5D SUSY theories have global ${\cal R}$-symmetries, some of which after
compactification survive in our $G_{44}$ model. These symmetries can be
successfully employed for lepton and baryon number conservations \cite{king},
\cite{our422}. 
Namely, terms in (\ref{yuk2}) have an accidental ${\cal R}$-symmetry:
$F\to e^{{\rm i}\alpha }F$, $F^c\to e^{-{\rm i}\alpha }F^c$. This symmetry
automatically forbids all matter parity violating couplings as well as
$d=5$ operators such as $FFFF$, $F^cF^cF^cF^c$. 
Thus, in this case, the LSP is expected to be stable.
Also, since gauge
mediated nucleon decay is absent in our model, conservation of baryon
number is guaranteed to all orders in perturbation theory.
Note that
lepton number is also conserved at this stage by the same 
${\cal R}$-symmetry\footnote{
For avoiding large Dirac type neutrino masses $\nu^clh_u$ from
(\ref{yuk2}), one can introduce an additional singlet state $N$ with a suitable
${\cal R}$ charge. Through the coupling 
$NF^c\ov{H}^{\hs{1mm}c}$ the state $\nu^c$ decouples
forming a massive ($\sim M_G$) state with $N$.}.

If we wish to accommodate the atmospheric and solar neutrino data 
\cite{atm}, \cite{sol}
we should generate non-zero but tiny neutrino masses. In particular,
lepton number must be
broken. It turns out that by modification of 
${\cal R}$-symmetry, one can violate lepton number, but still conserve B.
The appropriate fields transform as $F^c\to e^{{\rm i}\al }F^c$, 
$F\to e^{3{\rm i}\al }F$, $h\to e^{{\rm i}\bt }h$,
$S\to e^{{\rm i}(4\al -\bt )}S$, 
$(\Si_{-2}, \Si_2)\to (\Si_{-2}, \Si_2)$, with the superpotential 
$W\to e^{4{\rm i}\al }W$. The relevant couplings are

\beq
FF^c\Si_{-2}+S\Si_2h~.
\la{relcoup}
\eeq
The MSSM doublet-antidoublet pair now resides in $\Si_{-2}$, while
$\Si_2$ decouples with $h$ [second term in (\ref{relcoup})] forming
mass $\sim \lan S\ran \sim M_G $.

Note also that $\tl{U}(1)\times {\cal R}$-symmetry guarantee a zero $\mu $
term, and its generation should be achieved by some additional mechanism
(one possibility is a non-minimal K\"ahler potential \cite{muterm}).

The terms in (\ref{supHD4d}) are consistent with this ${\cal R}$-symmetry,
with
transformation properties 
$(H^c, \ov{H}^{\hs{1mm}c})\to e^{{\rm i}\al }(H^c, \ov{H}^{\hs{1mm}c})$,
$D_{1, 2}\to e^{2{\rm i}\al }D_{1, 2}$. Note that all the D and F-terms in
(\ref{lagf}) are still allowed by 
${\cal R}$-symmetry with
$\ov{F}\to e^{{\rm i}\al }\ov{F}$, 
$\ov{F}^c\to e^{4{\rm i}\al } \ov{F}^c$,
$\Psi \to e^{{\rm i}\bt }\Psi $ (e.g. all fragments living in $\Psi $),
$\ov{\Psi }\to e^{{\rm i}(4\al -\bt ) }\ov{\Psi }$, 
$(H, \ov{H})\to e^{{\rm i}\al }(H, \ov{H})$,
$(\Om , \ov{\Om })\to e^{2{\rm i}\al }(\Om , \ov{\Om })$. Thus, 
${\cal R}$-symmetry applies to the full theory. Since the 
${\cal R}$-charges of $H^c$ and $F^c$ superfields are the same,
we must impose 'matter' parity by hand in order to eliminate
some undesirable couplings.

The couplings generating Majorana masses for the right handed neutrinos
(in $F^c$), read

\beq
\fr{1}{M_P}(F^c\ov{H}^{\hs{1mm}c})^2~.
\la{maj}
\eeq
{}From (\ref{maj}), $M_R\sim M_G^2/M_P$ and neutrino mass
$m_{\nu }\sim \fr{h_u^2}{M_G^2}M_P\sim 0.1$~eV is readily obtained
(we have taken $M_P=2.4\cdot 10^{18}$~GeV, the reduced Planck mass,
and $h_u$ denotes the VEV of the higgs doublet that gives rise to Dirac
neutrino mass), just the scale
needed for explaining the atmospheric anomaly.
The scale for solar neutrinos can be obtained through a suppression of
appropriate Yukawa couplings in the Dirac type neutrino mass matrix. For
realizing desirable values for mixing angles within various oscillation
scenarios, the mechanisms suggested in \cite{our422}, \cite{osc} can be
applied.

Turning to the issue of baryon number conservation, the Planck scale
$d=5$ operators $\fr{1}{M_P}(FFFF+F^cF^cF^cF^c)$ are forbidden by 
${\cal R}$-symmetry. As far as the couplings 
$qqT+ql\bar T+u^cd^c\bar T+u^ce^cT$ are concerned 
($T$, $\bar T$ indicate colored
triplet states which could induce $d=5$ operators, after they are
integrated out), due to ${\cal R}$ charge prescriptions the couplings 
$qqT+ql\bar T$ do not emerge at all. 
The coupling $u^ce^c\ov d_{\ov H^c}^{\hs{1mm}c}$ emerges from (\ref{maj})
after extracting from $\ov{H}^{\hs{1mm}c}$ the triplet state
$\ov d_{\ov H^c}^{\hs{1mm}c}$.
The coupling $F^cF^cD_2$ yields
$u^cd^cd_{D_2}^c$. However, looking at (\ref{tripmas}), we see
that
there is no mass insertion between
$\ov d_{\ov H^c}^{\hs{1mm}c}$ and $d_{D_2}^c$ states, so the appropriate $d=5$
operators do not emerge. This means that baryon number is conserved in
our model.

\section{Conclusions}

In conclusion, we note that the gauge group $G_{44}$ can itself be
embedded in
a higher dimensional $SO(12)$ model. The decomposition of appropriate $SO(12)$
representations in
terms of $G_{44}$ are as follows:

$$
12=(6,~ 1)+(1,~ 6)~,~~~~~32=(4,~ 4)+(\bar 4,~ \bar 4)~,~~
$$
\beq
66=(6,~ 6)+(15,~ 1)+(1,~ 15)~,
\la{decso12}
\eeq 
We see on the right hand sides
all of the $G_{44}$ multiplets involved in our 5D SUSY
$G_{44}$ scenario.
Therefore, it is reasonable to think about higher dimensional 
(say ${\rm D}=6$)
unification of $G_{44}$ in $SO(12)$. The breaking of $SO(12)$
could occur through the steps
$SO(12)\to G_{44}\to G_{422}\times \tl{U}(1)$.
Details of these and related issues and their phenomenological
implications will be presented elsewhere.

\begin{acknowledgments} 
Q.S. would like to acknowledge the hospitality of  
the Alexander von Humboldt Stiftung and the Theory Group at DESY, 
especially Wilfried Buchm\"uller, while this work was in progress.
This work was supported in part by the DOE under Grant No. 
DE-FG02-91ER40626 
and by Nato under Grant No. PST.CLG.977666. 
\end{acknowledgments}

\bibliography{bl7_prd}% Produces the bibliography via BibTeX.

\end{document}